\documentclass[aps,pra,twocolumn,showpacs,groupedaddress]{revtex4}
\usepackage{epsfig}
\usepackage{mathrsfs}
\usepackage{subfigure}
\usepackage{graphicx}
\usepackage{indentfirst}
\usepackage{mathrsfs}
\usepackage{amsmath}
\usepackage{amssymb}
\usepackage{graphicx}
\usepackage{amsthm}
\usepackage[toc,page,title,titletoc,header]{appendix}
\usepackage{graphicx}
\usepackage{epstopdf}
\usepackage{CJK}
\theoremstyle{plain}

\theoremstyle{definition}

\theoremstyle{remark}

\begin{document}

\title{The charged Higgs bosons from the 3-3-1 models and the $\mathcal{R}(D^{(*)})$ anomalies.}

\author{Ma Wei}
\email{mawei@dlnu.edu.cn}
\affiliation{School of Physics and Materials Engineering, Dalian Nationalities University, Dalian 116600 China}

\author{Yue Chong-Xing}
\email{cxyue@lnnu.edu.cn}
\affiliation{Department of Physics, Liaoning Normal University, Dalian 116029, China}

\begin{abstract}
The several anomalies in the semileptonic B meson decays such as $\mathcal{R}(D^{(*)})$ have been reported by BaBar, Belle and LHCb collaborations recently. In this paper, we investigate the contributions of the charged Higgs bosons from the 3-3-1 models to the $\mathcal{R}(D^{(*)})$ anomalies. We find that, in a wide range of parameter space, the 3-3-1 models might give reasonable explanations to the $\mathcal{R}(D^{(*)})$ anomalies and other analogous anomalies of the B mesons semileptonic decays.
\end{abstract}

 \maketitle1

\section{Introduction}

In the recent years, the discovery of heavy quark spin-flavor symmetry and the formulation of heavy quark effective theory (HQET) [1] has shown that physical observable in the branching fractions of the semileptonic decays $B\rightarrow {D}^{(*)}l{\nu }_{l}$, which has drawn a lot of attentions, could be rather reliable predicted within the Standard Model (SM). Especially at the zero recoil point, which may allow a reliable determination of the Cabibbo-Kobayashi-Maskawa (CKM) element $V_{cb}$ [2]. In the SM, semileptonic decays of B mesons proceed via first-order electroweak interactions and are mediated by the W boson [3].  Since the effect of New Physics (NP) beyond the SM for these decays is induced by tree-level charged current, it is considered to be tiny.

There is an interesting phenomenon worth to study, the expression can be written as
$$\left.\mathcal{R}({D}^{(*)})=\frac{\mathcal{B}(\bar{B}\rightarrow {D}^{(*)}{\tau }^{-}{\bar{\nu}}_{\tau } )}{\mathcal{B}(\bar{B}\rightarrow {D}^{(*)}{l}^{-}{\bar{\nu}}_{l} )}\right |_{l\in\{e,\mu\}},$$
where the SM predictions are given by $\mathcal{R}(D)_{SM}=0.300\pm0.008$ [4] and $\mathcal{R}(D^{*})_{SM}=0.252\pm0.003$ [5], respectively. However, the present experimental values measured by BaBar [6], Belle [7] and LHCb [8] collaborations have recently observed anomalies in the ratios $\mathcal{R}(D)$ and $\mathcal{R}(D^{*})$. The average values given by the Heavy Flavor Average Group (HFAG) are $\mathcal{R}(D)_{avg}=0.397\pm0.040\pm0.028$ and $\mathcal{R}(D^{*})_{avg}=0.316\pm0.016\pm0.010$ [9], which exceed the SM predictions by 1.9$\sigma$ and 3.3$\sigma$, respectively. If one takes into account the $\mathcal{R}(D)$-$\mathcal{R}(D^{*})$ correlation of -0.21, the tension with the SM expectation would be at 4$\sigma$ level [10]. The anomalies caused wide concerns, that many works have been done, some of which were within model-independent frameworks [11-15], others were in NP models. The works in NP models are classified as mediated by leptoquarks [2,11,12,16,17], charged Higgs bosons [10,17,18], charged vector bosons [11, 20], and sparticles [21].

So far, the measurements of the ratios $\mathcal{R}(D^{(*)})$ in both the BABAR and Belle experiments used only the leptonic channels for the identification of $\tau$. When one  measures the branching fractions of the purely leptonic decay $B^{+}\rightarrow\tau^{+}\nu_{\tau}$, it will use both the leptonic and hadronic channels. Therefore, one kind of new ratios $\mathcal{R}_{\tau}(D^{(*)})$ are introduced, which are defined as
\begin{equation}\label{1}
\mathcal{R}_{\tau}(D^{(*)})=\frac{\mathcal{R}(D^{(*)})}{\mathcal{B}(B^{+}\rightarrow\tau^{+}\nu_{\tau})},
\end{equation}
then the systematics of $\tau$ detection tend to cancel, which provide us a more reliable test of the SM [26]. In the literature [2], the authors calculated the ratios $\mathcal{R}$ and $\mathcal{R}_{\tau}$ for the semileptonic decays $B\rightarrow D^{(*)}\tau\nu$, $B\rightarrow X_{c}\tau\nu$ and $B\rightarrow\pi\tau\nu$, which are induced by charged Higgs boson both in the SM and 2HDM-II. With measurement of $B\rightarrow \tau\nu$ and refined measurements of observable in $B\rightarrow D\tau\nu$ in different experiments, the study of $\mathcal{R}(D^{(*)})$ anomalies will be an effective solution for us to explore the SM and NP models. On the other hand, NP models might give rational interpretation for the $\mathcal{R}(D^{(*)})$ anomalies.

Among the new physics models, the so called 3-3-1 models [22-28] with gauge symmetry $SU(3)_{C}\times SU(3)_{L}\times U(1)_{X}$ are interesting extension of the SM, which can explain why there are three family fermions and why there is quantization of electric charge. In the general 3-3-1 models, there are $3\times 3\times 2=18$ scalar states, namely, four pairs of singly-charged states, five CP-odd states and five CP-even states. After some of the states being absorbed into the gauge bosons, we have the physical singly-charged Higgs bosons $H^{\pm}$, the CP-odd Higgs boson $A$ and the CP-even Higgs boson $H_{a}$. With the additional scalar bosons above, the 3-3-1 models may give us rich phenomenology. So far, the extended Higgs sectors have not yet been ruled out experimentally, and there have been already many works on the study of the Higgs bosons from the 3-3-1 models [29]. Since the charged Higgs bosons $H^{\pm}$ can couple to the SM quarks and leptons, we reasonably infer that the 3-3-1 models may give explanations to the $\mathcal{R}(D^{(*)})$ anomalies.

This paper is organized as follows. In Sec.2, we briefly review the essential features of the 3-3-1 models. The relevant couplings of the charged Higgs bosons to other particles are also discussed in this section. In Sec.3, we use the latest related experimental data to give constraints on the parameters of the 3-3-1 models. To make the study more comprehensive, we also consider another two useful semileptonic decay modes $B\rightarrow X_{c}\tau\nu_{\tau}$ and $B\rightarrow\pi\tau\nu_{\tau}$ by the similar methods. Our conclusions are given in Sec.4.

\section{The basics content of the 3-3-1 models.}\label{TDFS}

The 3-3-1 models are based on the gauge symmetry $SU(3)_{C}\times SU(3)_{L}\times U(1)_{X}$, in which the electric charge operator is defined as
\begin{equation}\label{Q}
Q=T_{3}+\beta T_{8}+XI,
\end{equation}
where $T_{3}$ and $T_{8}$ are two of eight generators of $SU(3)_{L}$, X is the new quantum number corresponding to $U(1)_{X}$, the free parameter $\beta$ defines the different particle structure and is used to label the particular type of the 3-3-1 models. To avoid the presence of exotic charges in the fermion and gauge boson sectors, one prefer to choose the scheme of $\beta=\pm1/\sqrt{3}$.

In the 3-3-1 models, symmetry breaking is generally accomplished in two steps
\begin{eqnarray*}
                 &&SU(3)_{C}\times SU(3)_{L}\times U(1)_{X} \\ \nonumber
                 &&\xrightarrow[ u ]{} SU(3)_{C}\times SU(2)_{L}\times U(1)_{X}\xrightarrow[v_{1}, v_{2}]{} SU(3)_{C}\times U(1)_{X}
\end{eqnarray*}
which can be realized via developing nonzero vacuum expectation values (VEVs) $u$, $v_{1}$ and $v_{2}$ for the neutral component fields of three triplets.

The quark content of the 3-3-1 models is generally described by
\begin{eqnarray*}
&&q_{mL}=\left(
         \begin{array}{c}
           u_{m} \\
           -d_{m} \\
           B_{m} \\
         \end{array}
       \right)_{L}\sim (3^{*}, 3, 0), \\ \nonumber
&&q_{3L}=\left(
         \begin{array}{c}
           u_{3} \\
           d_{3} \\
           T_{3} \\
         \end{array}
       \right)_{L}\sim (3, 3, 1/3), \\ \nonumber
&&d^{c}\sim (3^{*}, 1, 1/3),\ u^{c}\sim (3^{*}, 1, -2/3),\ \\ \nonumber
&&T^{c}\sim (3^{*}, 1, -2/3),\ B^{c}_{m}\sim (3^{*}, 1, 1/3),
\end{eqnarray*}
where $m=1,2$ and the numbers in the parenthesis express their assigned quantum numbers of the group $SU(3)_{C}\times SU(3)_{L}\times U(1)_{X}$.

For the leptonic sector, each lepton family is arranged in triplets: the first two elements are the charged and neutral lepton, and the third element is a conjugate of the charged lepton or the neutral lepton. There are
\begin{eqnarray*}
&&\Psi_{nL}=\left(
            \begin{array}{c}
              e^{-}_{n} \\
              \nu_{n} \\
              N_{n}^{0} \\
            \end{array}
          \right)_{L} \sim (1, 3^{*}, -1/3), \\ \nonumber
&&\Psi_{L}=\left(
           \begin{array}{c}
             \nu_{1} \\
             e_{1}^{-} \\
             E_{1}^{-} \\
           \end{array}
         \right)_{L}\sim (1, 3, -2/3),\\ \nonumber
&&\Psi_{4L}=\left(
            \begin{array}{c}
              E_{2}^{-} \\
              N_{3}^{0} \\
              N_{4}^{0} \\
            \end{array}
          \right)_{L}\sim (1, 3^{*}, -1/3), \\ \nonumber
&&\Psi_{5L}=\left(
            \begin{array}{c}
              N_{5}^{0} \\
              E_{2}^{+} \\
              e_{3}^{+} \\
            \end{array}
          \right)_{L}\sim (1, 3^{*}, 2/3), \\ \nonumber
&&e^{c}_{n}\sim (1, 1, 1), \quad e_{3}^{c}\sim(1, 1, 1), \\ \nonumber
&&E_{1}^{c}\sim (1, 1, 1), \quad E_{2}^{c}\sim (1, 1, 1),
\end{eqnarray*}
where $n=2, 3$. For this kind of 3-3-1 models, besides the ordinary gauge bosons $\gamma$, $Z$ and $W^{\pm}$, new neutral gauge boson $Z'$, charged and neutral bileptons $V^{\pm}$ and $X^{0}$ are predicted.

The scalar fields of the 3-3-1 models are generally parameterized as
\begin{eqnarray}
  \Phi_{1} = \left(
                  \begin{array}{c}
                    \phi_{1}^{0} \\
                    \phi_{1}^{+} \\
                    \eta_{1}^{+} \\
                  \end{array}
                \right),
   \quad \Phi_{2}=\left(
           \begin{array}{c}
             \phi_{2}^{-} \\
             \phi_{2}^{0} \\
             \eta^{0}_{2} \\
           \end{array}
         \right),
   \quad \varphi=\left(
           \begin{array}{c}
             \eta_{3}^{-} \\
             \eta^{0}_{3} \\
             \phi_{3}^{0} \\
           \end{array}
         \right),
\end{eqnarray}
which correspond to $3\times 3\times 2=18$ scalar states, namely, four pairs of singly-charged states, five CP-odd states and five CP-even states. After eight states of which being absorbed into the longitudinal components of two pairs of charged gauge bosons ($W$ and $W'$) and four neutral gauge bosons ($Z$, $Z'$, $Y_{1}$ and $Y_{2}$), we have the physical singly-charged Higgs bosons $H^{\pm}$, the CP-odd Higgs boson $A$ and the CP-even Higgs boson $H_{a}$ [26]. The contributions from the 3-3-1 models to the $\mathcal{R}(D^{(*)})$ and $\mathcal{R}_{\tau}(D^{(*)})$ anomalies may induced both by the charged Higgs bosons $H^{\pm}$ and the new gauge boson $W'$. However, the mass of the new gauge boson $W'$ is big, and the interactions between $W'$ and the SM particles are faint. Comparing with the contribution from the SM gauge boson W, the contribution from $W'$ is tiny enough to be ignored. Therefore, we will concentrate on calculating the contributions from the charged Higgs bosons $H^{\pm}$ in the following parts.

The Yukawa coupling for the charged Higgs boson $H^{+}$ from the 3-3-1 models is generally given by [26,27]
\begin{eqnarray}
  G_{\bar{u}dH^{+}}&=&-\frac{g}{2\sqrt{2}m_{W}}\left[V_{CKM}-V_{L}^{u\dag}\Delta V_{L}^{d}\right]^{*}_{ud}[m_{d}(1 \nonumber\\
 &&-\gamma_{5})tan\beta+m_{u}(1+\gamma_{5})cot\beta],
\end{eqnarray}
where $u$ and $d$ refer to up- and down-type quarks, $m_{W}$, $m_{u}$ and $m_{d}$ express the masses of the SM particles $W$, $u$ and $d$, $\Delta=diag(0,0,1)$. $V_{L}^{u}$ and $V_{L}^{d}$ are the unitary matrices of the left-handed quarks, which are given by [26,27]
\begin{eqnarray}
V_{L}^{u}=\left(
                 \begin{array}{ccc}
                   0.975 & -0.223 & 1.86\times10^{-3} \\
                   0.222 & 0.974 & 0.0518 \\
                   -0.01340 & -0.0501 & 0.999\\
                 \end{array}
               \right), \\
V_{L}^{d}=\left(
                 \begin{array}{ccc}
                   1.00 & 2.56\times10^{-3} & 5.87\times10^{-3} \\
                   -3.10\times10^{-3} & 0.996 & 0.0941 \\
                   -5.61\times10^{-3} & -0.0942 & 0.996 \\
                 \end{array}
               \right),
\end{eqnarray}
\textbf{where we have ignored CP violation, by which the experimental values of the elements of Cabibbo-Kobayashi-Maskawa(CKM) matrix defined as $V_{L}^{u}(V_{L}^{d})$ are reproduced. For the most important factor in Eq.(4) is $V_{CKM}$ matrix, the numerical results are not much changed if different $V_{L}^{u}$ and $V_{L}^{d}$ are used. Besides, one can see from Eq.(4) that when the mass of the up-type quark is much smaller than that of the down-type quark, the main contribution of the coupling is coming from the down-type quark term which is containing the parameter $tan\beta$. Which applies to the decay $b\rightarrow c\tau\nu$. Namely, the main contribution of the coupling is coming from b quark term, since $m_{c}< m_{b}$.}
The decay rates for the charged Higgs bosons $H^{\pm}$ from the 3-3-1 models into the SM charged lepton and neutrino pairs are generally give by
\begin{eqnarray}
 \Gamma(H^{+}\rightarrow l^{+}\nu)=\frac{G_{F}m_{H^{\pm}}}{4\sqrt{2}\pi}m_{l}^{2}tan^{2}\beta\left(1-\frac{m_{l}^{2}}{m^{2}_{H^{\pm}}}\right)^{2},
\end{eqnarray}
where $m_{l}$ denotes the mass of SM charged lepton, $m_{H^{\pm}}$ denote the masses of the charged Higgs bosons $H^{\pm}$.
The squared masses of $H^{\pm}$ are given by
$m_{H^{\pm}}^{2}=\frac{v^{2}}{2}\lambda_{4}+M^{2},$
where $v=\sqrt{v_{1}^{2}+v_{2}^{2}}=(\sqrt{2}G_{F})^{-1/2}\simeq246GeV$ with $G_{F}$ being the Fermi constant, $\lambda_{4}$ is one of the group parameters of the $SU(3)$ group. $tan \beta=v_{2}/v_{1}$ is a scalar mixing angle, which is possible to obtain a natural fit for the observed neutrino hierarchical and mixing angles when $tan \beta\gg \mathcal{O}(1)$[30]. $M^{2}=\mu u/(\sqrt{2}s_{\beta}c_{\beta})$, $s_{\beta}$ and $c_{\beta}$ are the shorthand notations for the angles $\beta$ (i.e. $c_{\beta}=cos \beta$, $s_{\beta}=sin \beta$) and $\mu$ is the soft breaking term involved in the Higgs potential. Since the charged Higgs bosons $H^{\pm}$ can couple to the SM quarks and leptons, we infer that the 3-3-1 models may give contributions to the $\mathcal{R}(D^{(*)})$ and $\mathcal{R}_{\tau}(D^{(*)})$ anomalies.

So far, the specific scope of masses of the charged Higgs bosons $H^{\pm}$ are not predicted by the 3-3-1 models.  Considering that there are no experimental upper bounds on the mass of the charged Higgs, one generally expects to have $m_{H}<1TeV$ in order to guarantee the perturbation theory remains valid [31]. In addition, the regions for $m_{H^{+}}\leq540GeV$ have already been excluded by $b\rightarrow s\gamma$ measurements at $95\%$ confidence level [32]. Therefore, in this paper, we will take $540GeV<m_{H^{+}}<1000GeV$ in the following numerical calculation.

\section{The contributions of the charged Higgs to the $\mathcal{R}(D^{(*)})$ and $\mathcal{R}_{\tau}(D^{(*)})$ anomalies.}\label{SNC}
The expression for the branching fraction $Br(B\rightarrow \tau\nu_{\tau})$ in the SM is given by
\begin{equation}\label{Brtaunu}
 \mathcal{B}^{SM}(B^{+}\rightarrow \tau^{+}\nu_{\tau})=\frac{G_{F}^{2}m_{B}m_{\tau}^{2}}{8\pi}\left[1-\frac{m_{\tau}^{2}}{m_{B}^{2}}\right]f_{B}^{2}\left|V_{ub}\right|^{2}\tau_{B^{+}},\nonumber
\end{equation}
where $\tau_{B^{+}}$ is the lifetime of the $B^{+}$ meson, $m_{B}$ and $m_{\tau}$ are the masses of $B$ meson and $\tau$ lepton. In our numerical calculation, we will take  $G_{F}=1.1663787(6)\times10^{-5} GeV^{-2}$, $m_{B}=5.27929\pm0.00015 GeV$, $m_{\tau}=1.77686\pm0.00012 GeV$, $m_{b}=4.20\pm0.07 GeV$, $f_{B}=0.191\pm0.007 GeV$, $\tau_{B^{+}}=1.638(4)ps^{-1}$, $|V_{ub}|=(3.61\pm0.32)\times10^{-3}$ [33-36].

In this section, we will calculate the contributions of the charged Higgs from the 3-3-1 models to the three kinds of decays $B\rightarrow D^{(*)}\tau\nu_{\tau}$, $B\rightarrow X_{c}\tau\nu_{\tau}$ and $B\rightarrow \pi\tau\nu_{\tau}$. One can see from relevant expressions that the parameters $tan \beta$ and $m_{H^{+}}$ are important, which may have great influence on the numerical results. Besides, considering the characteristics of the form of the two parameters in the related calculation formula, we define a new parameter $r=tan \beta/m_{H^{+}}$, which may reflect the synergy of the two parameters.
\subsection{$\mathcal{R}(D^{(*)})$ and $\mathcal{R}_{\tau}(D^{(*)})$}
The contributions from the 3-3-1 models to the $\mathcal{R}^{331}(D^{(*)})$ and $\mathcal{R}_{\tau}^{331}(D^{(*)})$ anomalies may induced both by the charged Higgs bosons $H^{\pm}$ and the new gauge boson $W'$. As we have discussed before, the contribution from the new gauge boson $W'$ is tiny enough to be ignored. Therefore, we only investigate the contributions from the charged Higgs bosons $H^{\pm}$ in the following parts.

After dropping the terms that are negligible, we have the compact expressions of $\mathcal{R}^{331}(D)$ and $\mathcal{R}^{331}(D^{*})$ ratios including the contributions of the 3-3-1 models, which can be approximately written as
\begin{eqnarray}
\mathcal{R}(D^{(*)})_{331}\!\approx\!\mathcal{R}(D^{(*)})_{SM}\!-\!A_{D^{(*)}}\!\frac{tan^{2}\!\beta}{m^{2}_{H^{+}}}\!+\!B_{D^{(*)}}\frac{tan^{4}\!\beta}{m^{4}_{H^{+}}},
 \end{eqnarray}
where $A(B)_{D^{(*)}}$ are coefficients determined by averaging over $B^{0}$ and $B$ deays [6]. $A_{D}=-3.25\pm0.32\ GeV^{2}$, $B_{D}=16.9\pm2.0\ GeV^{4}$, $A_{D^{*}}=-0.230\pm0.029\ GeV^{2} $, $B_{D^{*}}=0.643\pm0.085\ GeV^{4}$, $\mathcal{R}_{\tau}^{SM}(D)=(3.136\pm0.628)\times10^{3}$ and $\mathcal{R}_{\tau}^{SM}(D^{*})=(2.661\pm0.512)\times10^{3}$  [2]. One can see from Eq.(8) that the contributions from the 3-3-1 models to $\mathcal{R}(D^{(*)})$ anomalies are approximately proportional to the model parameter $tan\beta$ (or $\frac{tan\beta}{m_{H^{+}}}$).

\textbf{The ratios $\mathcal{R}^{331}_{\tau}(D^{(*)})$ can be written as}
\begin{eqnarray}
\mathcal{R}^{331}_{\tau}(D^{(*)})&=&\frac{\mathcal{R}^{331}(D^{(*)})}{\mathcal{B}(B^{+}\rightarrow\tau^{+}\nu_{\tau})}\nonumber\\
&\approx&\mathcal{R}^{\!S\!M}_{\tau}(D^{(*)})-\!\frac{A_{D^{(*)}}}{\mathcal{B}(\!B^{+}\!\rightarrow\!\tau^{+}\!\nu_{\tau}\!)}\cdot\frac{tan^{2}\beta}{m^{2}_{H^{+}}}\nonumber\\
&&+\frac{B_{D^{(*)}}}{\mathcal{B}(B^{+}\!\rightarrow\!\tau^{+}\!\nu_{\tau}\!)}\cdot\frac{tan^{4}\beta}{m^{4}_{H^{+}}},
\end{eqnarray}
with
\begin{eqnarray}
\mathcal{B}(\!B^{+}\!\rightarrow\!\tau^{+}\!\nu_{\tau}\!)_{331}&=&\mathcal{B}(B^{+}\!\rightarrow\!\tau^{+}\!\nu_{\tau}\!)_{SM}[1+(V_{CKM}\nonumber \\
                                               &&\left.\left.-V_{L}^{u\dag}\Delta V_{L}^{d}\right)^{2}_{13} \frac{m_{B}^{2}}{V_{ub}^{2}}\frac{tan^{2}\beta}{m_{H^{+}}^{2}}\right]^{2}\!.
\end{eqnarray}
\textbf{The values of all the parameters in Eq.(9)-Eq.(10)} have been given, which are not listed again.

 The latest measured values of $\mathcal{R}(D^{(*)})$, $\mathcal{R}_{\tau}(D^{(*)})$ and $\mathcal{B}(B^{+}\rightarrow\tau^{+}\nu_{\tau})$ by BABAR and Belle are summarized in Tab.I [2, 6, 7]:
\begin{table}[htbp]
\centering  
\begin{tabular}{|c|c|c|}  
\hline
                                                          & BABAR                          & Belle                       \\  \hline  
$\mathcal{R}(D)$                                          & $0.440\pm0.058\pm0.042$        & $0.375\pm0.064\pm0.026$     \\  \hline 
$\mathcal{R}(D^{*})$                                      & $0.332\pm0.024\pm0.018$        & $0.302\pm0.030\pm0.011$     \\  \hline
$\mathcal{B}(B^{\!+\!}\!\rightarrow\!\tau^{\!+\!}\!\nu_{\!\tau\!})$        & $1.83^{+0.53}_{-0.49}\times10^{-4}$   & $(1.25\pm0.28)\times10^{-4}$    \\  \hline
$\mathcal{R}_{\tau}(D)$                                   & $(2.404\pm0.838)\times10^{3}$& $(3.0\pm1.1)\times10^{3}$       \\  \hline
$\mathcal{R}_{\tau}(D^{*})$                               & $(1.814\pm0.5282)\times10^{3}$& $(2.416\pm0.794)\times10^{3}$   \\  \hline
\end{tabular}
\caption{The measured values of $\mathcal{R}(D)$, $\mathcal{R}(D^{*})$ and $\mathcal{B}(B^{+}\rightarrow\tau^{+}\nu_{\tau})$ by BABAR and Belle.}
\end{table}

\textbf{One can see from Eq.(8)-Eq.(10)}, that the values of the rates $\mathcal{R}(D^{(*)})$ and $\mathcal{R}_{\tau}(D^{(*)})$ depend on the model parameters $m_{H^{+}}$ and $tan\beta$. We use the above experimental data to bound the parameters of the 3-3-1 models, and find that the reasonable parameter values can explain the $\mathcal{R}(D^{(*)})$ anomalies. In order to see the allowed parameter space of the 3-3-1 models by BABAR and Belle data, we plot $tan\beta$ as a function of $m_{H^{+}}$ in Fig.1a (left) and Fig.1b (right), respectively. One can see from Fig.1a that the largest parameter space is blue region, which is coming from $\mathcal{R}(D)$ data. The smallest region also denotes the common allowed parameter space for the three data sets, is yellow region, which is coming from $\mathcal{B}(B^{+}\rightarrow\tau^{+}\nu_{\tau})$ data. We found that the constraint on the parameter space from $\mathcal{B}(B^{+}\rightarrow\tau^{+}\nu_{\tau})$ data is the biggest. In the common space, the reasonable value of $tan\beta$ is in the range of $27\sim145$. The conclusions of Fig.1b are similar to those of Fig.1a, and the reasonable value of $tan\beta$ is in the range of $0\sim108$ by Belle data.

\begin{figure}[htbp]
\centering
\includegraphics[width=121pt,height=120pt]{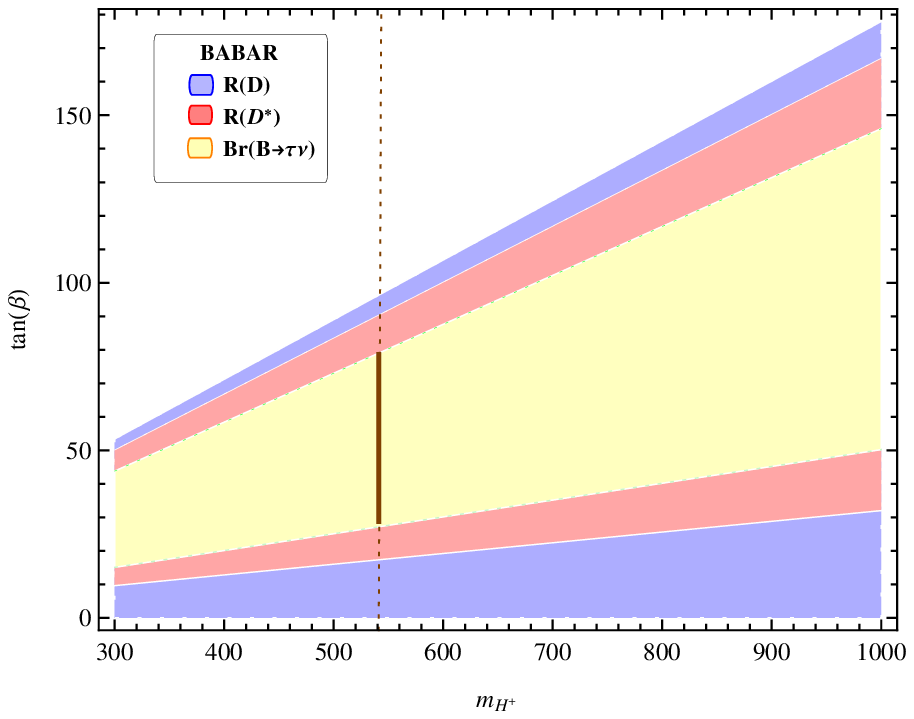}
\includegraphics[width=121pt,height=120pt]{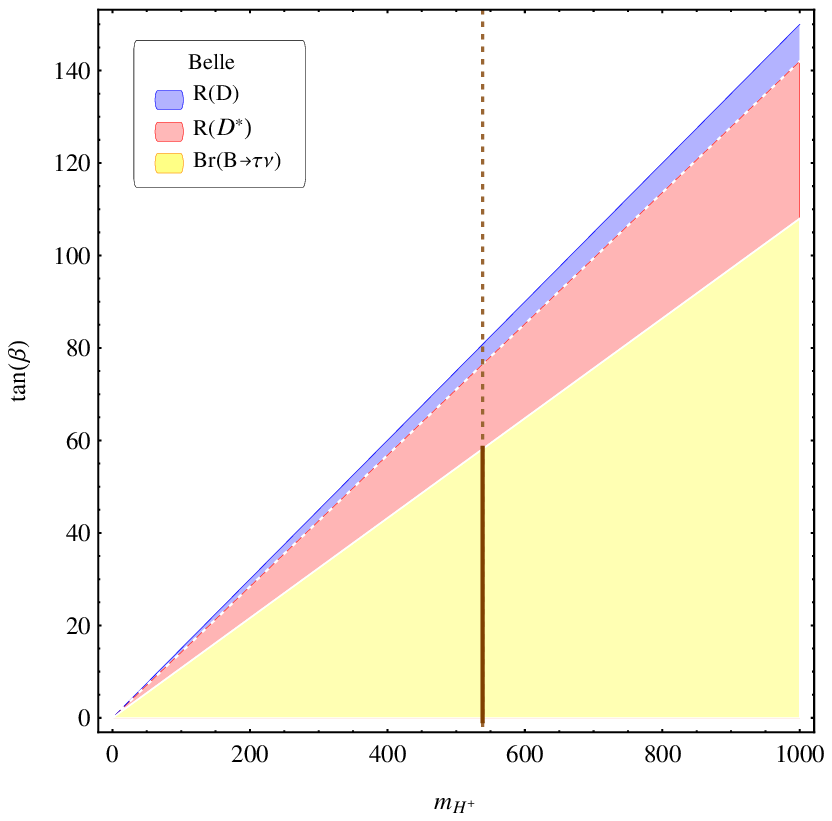}
\caption{For the 3-3-1 models, the allowed parameter spaces by BABAR (left) and Belle (right) data for $\mathcal{R}(D)$, $\mathcal{R}(D^{*})$ and $\mathcal{B}(B^{+}\rightarrow\tau^{+}\nu_{\tau})$. The dotted vertical lines show $m_{H^{+}}= 540 GeV$.}
\end{figure}
In order to see the synergistic effect of the parameters $tan \beta$ and $m_{H^{+}}$ on the ratios $\mathcal{R}_{\tau}(D^{(*)})$, we plot $\mathcal{R}_{\tau}(D^{(*)})$ as a function of parameter $r=tan \beta/m_{H^{+}}$ in Fig.2. In our numerical calculation, we assume $540GeV\leq m_{H^{+}}\leq1000GeV$ for three typical values of $tan\beta$ ($tan\beta=2,10,20$). We can see from Fig.2 that it is possible for the 3-3-1 models to explain the $\mathcal{R}_{\tau}(D^{(*)})$ anomalies when $0<tan\beta\leq10$.
\begin{figure}[htb]
\centering
\includegraphics[width=120pt,height=120pt]{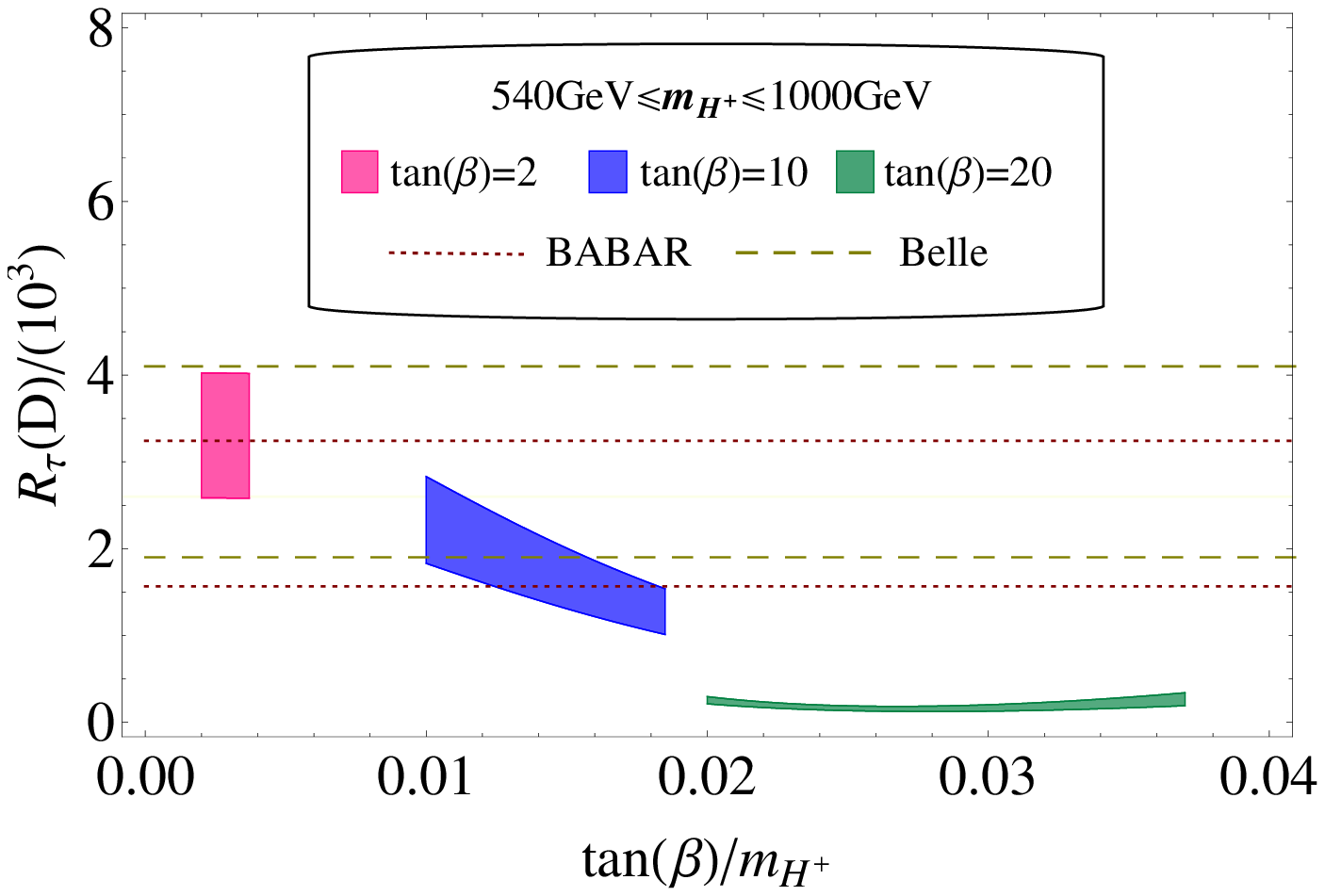}
\includegraphics[width=120pt,height=120pt]{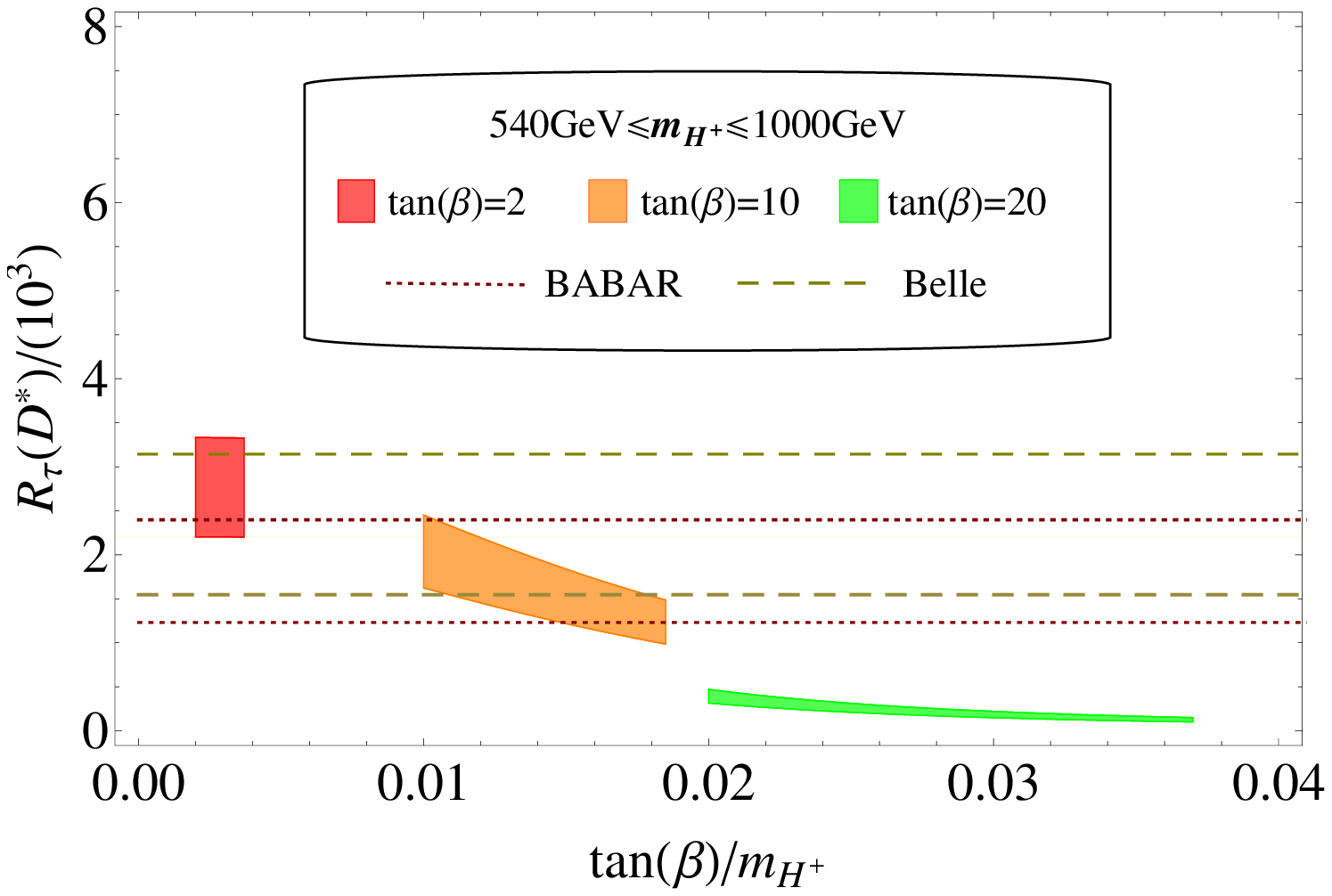}
\caption{Variations of $\mathcal{R}_{\tau}(D)$ (left) and $\mathcal{R}_{\tau}(D^{*})$ (right) with the parameter $r=tan \beta/m_{H^{+}}$ for three dispersive
values of $tan \beta$. The 1$\sigma$ experimental ranges given by BABAR and Belle are shown as dotted and dashed horizontal lines.}
\end{figure}

\subsection{$\mathcal{R}(X_{c})$ and $\mathcal{R}_{\tau}(X_{c})$}
In this subsection, we discuss the inclusive semileptonic decay channels of the B meson $B\rightarrow X_{c}\tau\nu_{\tau}$. The ratios $\mathcal{R}(X_{c})$ and $\mathcal{R}_{\tau}(X_{c})$ can be defined as $$\mathcal{R}(X_{c})=\frac{\mathcal{B}(B\rightarrow X_{c}\tau\bar{\nu})}{\mathcal{B}(B\rightarrow X_{c}e\bar{\nu})}, \quad \mathcal{R}_{\tau}(X_{c})=\frac{\mathcal{R}(X_{c})}{\mathcal{B}(B^{+}\rightarrow\tau^{+}\nu_{\tau})}.$$
The SM prediction and experimental measurement values for $\mathcal{R}(X_{c})$ and $\mathcal{R}_{\tau}(X_{c})$ are summarized in Tab.II [26, 37-39]. One can see from Tab.II that there are also similar anomalies in the $\mathcal{R}(X_{c})$ and $\mathcal{R}_{\tau}(X_{c})$ ratios. Thus, we calculate the contributions of the charged Higgs from the 3-3-1 models to the $\mathcal{R}(X_{c})$ anomaly along with the $\mathcal{R}(D^{(*)})$ anomalies.
\begin{table}[htbp]
\centering  
\begin{tabular}{|l|c|c|c|}  
\hline
      & $\mathcal{R}(X_{c})$   &  $\mathcal{B}(B^{\!+\!}\!\rightarrow\!\tau^{\!+\!}\!\nu_{\!\tau\!})$ $\!(\!\times\!10^{\!-4}\!)$  &$\mathcal{R}_{\tau}(X_{c})$ $\!(\times\!10^{\!3\!})$                            \\ \hline  
SM    & $0.225\pm0.006$        &  $0.947\pm0.182$                        &$2.060^{+0.087}_{-0.083}$                 \\  \hline   
B\!A\!B\!A\!R & $0.221\pm0.021$        &  $1.83^{+0.53}_{-0.49}$                    &$1.208^{+0.598}_{-0.361}$                 \\  \hline              
Belle & $0.221\pm0.021$        &  $1.25\pm0.28$                          &$1.768^{+0.727}_{-0.461}$                 \\  \hline              
\end{tabular}
 \caption{The SM prediction and experimental measurement values for $\mathcal{R}(X_{c})$ and $\mathcal{R}_{\tau}(X_{c})$.}
\end{table}

\begin{figure}[htb]
\begin{center}
\epsfig{file=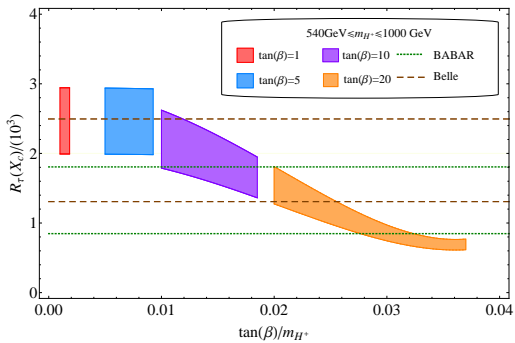,width=0.4\textwidth}
\caption{Variations of $\mathcal{R}_{\tau}(X_{c})$ with the parameter $r=tan \beta/m_{H^{+}}$ for four typical values of $tan \beta$. The 1$\sigma$ experimental ranges are shown by the dotted (BABAR) and dashed (Belle) horizontal lines.} \label{ee}
\end{center}
\end{figure}

The expressions of the differential decay rate of inclusive channel $B\rightarrow X_{c}\tau\nu_{\tau}$ have been given by HQET in context of the SM [40]. Considering the contributions of the 3-3-1 models, this expression forms can be similarly given. To save the length of the paper, we will not present them here. We plot $\mathcal{R}^{331}_{\tau}(X_{c})$ as a function of the parameter $r=tan \beta/m_{H^{+}}$ for decentralized values of $tan\beta$ and $540GeV\leq m_{H^{+}}\leq1000GeV$ in Fig.3. On account of the discussions above, we do not choose big values for $tan\beta$ here. Our numerical results show that for $0<tan\beta\lesssim20$ and $540GeV\leq m_{H^{+}}\leq1000GeV$, the 3-3-1 models might give reasonable explanation to the $\mathcal{R}_{\tau}(X_{c})$ anomaly.

\begin{figure}[htb]
\begin{center}
\epsfig{file=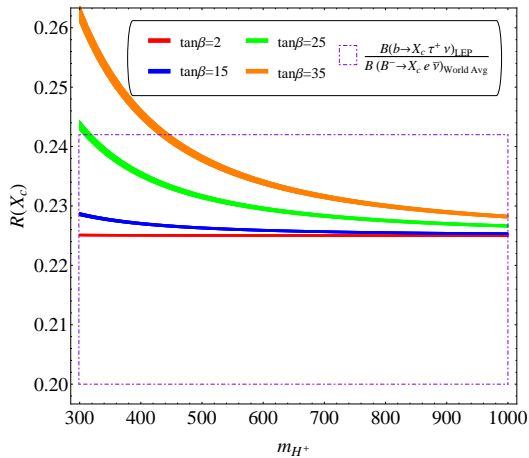,width=0.4\textwidth}
\caption{Variations of $\mathcal{R}(X_{c})$ with the mass parameter $m_{H^{+}}$ for four typical $tan \beta$ values. Experimental range is
shown by violet dot-dashed lines. } \label{ee}
\end{center}
\end{figure}

In order to see the effect of the model parameter $m_{H^{+}}$ on the ratio $\mathcal{R}(X_{c})$, we plot $\mathcal{R}(X_{c})$ as a function of $m_{H^{+}}$ for four dispersive values of the parameter $tan\beta$ in Fig.4. The colorized thick lines actually denote corresponding range of values and the violet dot-dashed lines denote the experimental range $\mathcal{R}(X_{c})_{Exp}$, which is given by $$\mathcal{R}(X_{c})_{Exp}=\frac{\mathcal{B}(b\rightarrow X_{c}\tau\bar{\nu})_{LEP}}{\mathcal{B}(B\rightarrow X_{c}e\bar{\nu})_{World\ Avg}},$$ where $\mathcal{B}(b\rightarrow X_{c}\tau\bar{\nu})_{LEP}=(2.41\pm0.23)\%$ and $\mathcal{B}(B\rightarrow X_{c}e\bar{\nu})_{World\ Avg}=(10.92\pm0.16)\%$. And we obtain $\mathcal{R}(X_{c})_{Exp}=0.22^{+0.025}_{-0.023}$. One can see from Fig.4 that $\mathcal{R}^{331}(X_{c})$ is sensitive to both the parameters $tan\beta$ and $m_{H^{+}}$. For $0\leqslant tan\beta\leqslant35$ and $540GeV\leqslant m_{H^{+}}\leqslant1000GeV$, the value of $\mathcal{R}(X_{c})$ is in a range of $0.225\sim0.2366$, where the value $0.225$ corresponds to the SM prediction.

\subsection{$\mathcal{R}(\pi)$ and $\mathcal{R}_{\tau}^{\pi}$}
In this subsection, we calculate the contributions of the charged Higgs bosons $H^{\pm}$ from the 3-3-1 models to the semileptonic decay channels of the B meson $B\rightarrow\pi\tau\nu_{\tau}$. The previous researches of this mode [41,42] were with the SM leptons $\mu$ and $e$ to extract the CKM element $V_{ub}$ [33]. The observable is defined as $$\left.\mathcal{R}(\pi)=\frac{\mathcal{B}(B\rightarrow\pi\tau\bar{\nu}_{\tau}))}{\mathcal{B}(B\rightarrow \pi l\bar{\nu}_{l})}\right|_{l\in\{e,\mu\}},$$ which is potentially sensitive to NP, where the dependence and the uncertainty due to $V_{ub}$ were canceled. In order to keep the dependence on $V_{ub}$ reserved, there is another kind of useful definition [26] $$\mathcal{R}_{\tau}^{\pi}=\frac{\mathcal{B}(B\rightarrow\pi\tau\bar{\nu}_{\tau})}{\mathcal{B}(B\rightarrow\tau\bar{\nu}_{\tau})},$$ which is made minor adjustments to the form of Eq.(1). We summarize the present relevant data in Tab.III [26, 41]. One can see from Tab.III that there are big deviation between the experimental data and the SM predictions. The method we use in this subsection, is the same as what we used when we studied $\mathcal{R}(D^{(*)})$ and $\mathcal{R}_{\tau}(D^{(*)})$.
\begin{table}[htbp]
\centering
\begin{tabular}{|l|c|c|}
\hline
    & $\mathcal{R}(\pi)$                      &  $\mathcal{R}_{\tau}^{\pi}$                   \\ \hline
SM  & $0.641\pm0.016$                     &  $0.733\pm0.144$                            \\ \hline
Exp & $<1.784$             &  $<2.62$                  \\  \hline
\end{tabular}
\caption{The values of $\mathcal{R}(\pi)$ and $\mathcal{R}_{\tau}^{\pi}$ coming from the SM prediction and the upper limits based on the current experiment average.}
\end{table}
\begin{figure}[htb]
\begin{center}
\epsfig{file=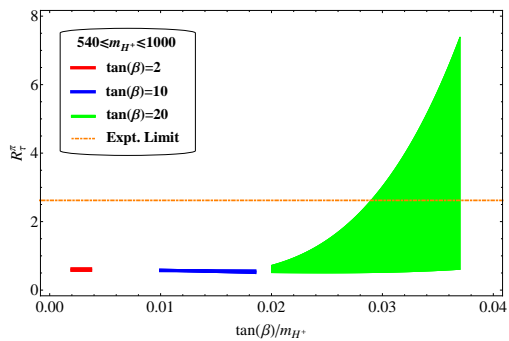,width=0.4\textwidth}
\caption{Variations of $\mathcal{R}_{\tau}^{\pi}$ as a function of the parameter $r=tan \beta/m_{H^{+}}$ for three dispersive
values of $tan \beta$ and $540 \leq m_{H^{+}} \leq 1000$. Experimental upper limit is shown by the orange dot-dashed horizontal line.} \label{ee}
\end{center}
\end{figure}

\begin{figure}[htb]
\begin{center}
\epsfig{file=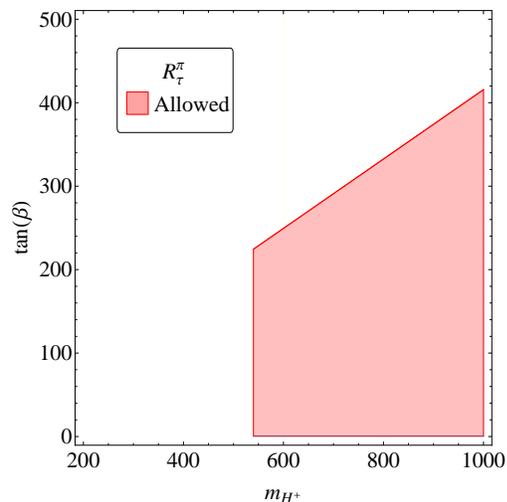,width=0.4\textwidth} \vspace{-0.6cm}
\caption{Allowed parameter space for $tan \beta$ and $m_{H^{+}}$ obtained from the analysis of $\mathcal{R}_{\tau}^{\pi}$ in the 3-3-1 models.} \label{ee}
\end{center}
\end{figure}

In Fig.5, we plot $\mathcal{R}_{\tau}^{\pi}$ as a function of the parameter $r=tan \beta/m_{H^{+}}$. We note that for $0<tan\beta\lesssim20$ and $540GeV\leq m_{H^{+}}\leq1000GeV$, the 3-3-1 models might give explanation to the $\mathcal{R}_{\tau}^{\pi}$ anomaly. We also plot the allowed parameter space for $tan\beta$ and $m_{H^{+}}$ obtained from the analysis of $\mathcal{R}_{\tau}^{\pi}$ in Fig.6.

\section{Conclusions}
The so called $\mathcal{R}(D^{(*)})$ anomalies cause wide attention recently. Many works have been done, some of which were within model-independent frameworks, others were in NP models. It indicate that NP beyond the SM might give reasonable explanations to the $\mathcal{R}(D^{(*)})$ anomalies and analogous anomalies of other B mesons semileptonic decays.

The 3-3-1 models predict the charged Higgs bosons $H^{\pm}$, which can couple to the SM particles. Therefore, we have the motivation to study the charged Higgs bosons from the 3-3-1 models to the $\mathcal{R}(D^{(*)})$ anomalies and other analogous anomalies of the B mesons semileptonic decays. Furthermore, to make the study more comprehensive, we also calculate the contributions from the 3-3-1 models to two uesful B mesons semileptonic decays $B\rightarrow X_{c}\tau\nu_{\tau}$ and $B\rightarrow\pi\tau\nu_{\tau}$.

First, we calculate the contributions of the charged Higgs bosons $H^{\pm}$ to the $\mathcal{R}_{(\tau)}(D^{(*)})$ anomalies. Our numerical results show that there are large common allowed parameter spaces for both BABAR and Belle data. The biggest constraint on the 3-3-1 models parameter space comes from $\mathcal{B}(B^{+}\rightarrow\tau^{+}\nu_{\tau})$ data. In the common spaces, the reasonable values of $tan\beta$ are in the ranges of $27\sim145$ and $0\sim108$ by BABAR and Belle data, respectively. For $0<tan\beta\leq10$ and $540GeV\leq m_{H^{+}}\leq1000GeV$, the 3-3-1 models might give reasonable explanation to the $\mathcal{R}^{331}_{\tau}(D^{(*)})$ anomalies.

Then we calculate the contributions to the $\mathcal{R}_{(\tau)}(X_{c})$ anomalies. Our numerical results show that $\mathcal{R}_{(\tau)}(X_{c})$ are sensitive to both the parameters $tan\beta$ and $m_{H^{+}}$. The 3-3-1 models might give reasonable explanation to the $\mathcal{R}_{(\tau)}(X_{c})$ anomalies in the model parameters range of $0<tan\beta\lesssim20$.

Finally, we  investigate the contributions to the $\mathcal{R}(\pi)$ and $\mathcal{R}_{\tau}^{\pi}$ anomalies. Our numerical results show that for $0<tan\beta\lesssim20$, the 3-3-1 models might give reasonable explanation to the  $\mathcal{R}(\pi)$ and $\mathcal{R}_{\tau}^{\pi}$ anomalies.

Combining the observations and our numerical results, we draw the following conclusions. In a wide range of parameter space, the 3-3-1 models may give reasonable explanations to the $\mathcal{R}(D^{(*)})$ anomalies and  other analogous anomalies of B mesons semileptonic decays. With measurement of $B\rightarrow \tau\nu$ and refined measurements of observable in $B\rightarrow D\tau\nu$ in different experiments, the study of $\mathcal{R}(D^{(*)})$ anomalies will be an effective solution for us to probe both the SM and NP models.

\section*{Acknowledgments}
This work was supported in part by the National Natural Science Foundation of
China under Grant No. 11275088, the Natural Science Foundation of the Liaoning Scientific Committee
(No. 2014020151).

\end{document}